\documentclass[a4paper,11pt]{article}
\usepackage[DIV12]{typearea}

\usepackage[font=small,labelfont=bf]{caption}
\usepackage{graphicx}
\usepackage{amsmath}
\usepackage{amssymb}
\usepackage{verbatim}
\usepackage{enumerate}
\usepackage{bm}
\usepackage{color}
\usepackage{ulem}
\usepackage{cite}
\usepackage{colortbl}
\definecolor{light-gray}{gray}{0.85}
\usepackage{multirow}
\begin{document}
\thispagestyle{empty}
\begin{center}

{\Large\bf 750 GeV Diphotons and Supersymmetric Grand Unification}\\[12mm]

{\large Hans Peter Nilles, Martin Wolfgang Winkler} 
\\[6mm]
{\it Bethe Center for Theoretical Physics\\
and\\
Physikalisches Institut der Universit\"at Bonn\\
Nussallee 12, 53115 Bonn, Germany
}
\vspace*{12mm}
\begin{abstract}We investigate the 750 GeV diphoton excess in terms of supersymmetric models which preserve grand unification in the ultraviolet. We show that minimal extensions of the MSSM by a singlet and a vector-like 5-plet or 10-plet of SU(5) can explain the observed signal while remaining perturbative up to the GUT scale. 
Different from previous analyses we rely on light sfermions in the loops which -- compared to the  analog non-supersymmetric models -- enhance the diphoton cross section by up to a factor of seven. While the resonance decay width is narrow, mass splitting of the scalar and pseudoscalar components may result in a double resonance. We perform a likelihood analysis on the ATLAS and CMS data to show that the significance of the diphoton excess increases from $3.3 \,\sigma$ (single narrow resonance) to $3.9\,\sigma$ for the double resonance. We also provide signal predictions in other diboson channels to be tested at LHC-13.
\end{abstract}
\end{center}
\clearpage

\section{Introduction}

An excess of diphotons at ATLAS~\cite{ATLAS:13tev} and CMS~\cite{CMS:2015dxe} has triggered speculations about the existence of a new boson in nature. If taken serious, the properties of this new particle were not foreseen in any preexisting model of particle physics. While one may take this as an argument for a statistical fluctuation, it is still tempting to ask, how the new boson can be embedded into well-motivated schemes of physics beyond the standard model. An obvious candidate for the diphoton excess is a singlet which couples to gluons (photons) via new vector-like colored (charged) particles running in the loop. Simple bottom-up constructions of this type were presented in~\cite{Angelescu:2015uiz,Knapen:2015dap,Ellis:2015oso,Falkowski:2015swt,Franceschini:2015kwy}. 
Vector-like `exotics' are very common in string model constructions such as D-brane theories~\cite{Dijkstra:2004cc} or heterotic string compactifications~\cite{Lebedev:2006kn}. The possibility that these exotics mediate the diphoton signal has been considered in~\cite{Heckman:2015kqk,Cvetic:2015vit,Cvetic:2016omj,Palti:2016kew,Karozas:2016hcp,Faraggi:2016xnm,Li:2016tqf,Leontaris:2016wsy}.
However, in UV derived settings, there exist strong constraints on the possible charges and couplings of the new states. In particular, it proofs very difficult to construct viable models which fit the signal and remain perturbative up to high energies.

In this article we consider supersymmetric extensions of the standard model consistent with grand unification. The most economic way to preserve gauge coupling unification is to introduce vector-like states as complete multiplets of an SU(5) grand unified theory (GUT). In this scenario, the couplings of the singlet resonance to vector-like multiplets are limited by the the absence of Landau poles in the renormalization group running. This results in an upper limit on the diphoton cross section of the 750 GeV boson. As a consequence, it was found in~\cite{Hall:2015xds} that simple realizations with one/ two $5+\overline{5}$ or one $10+\overline{10}$ and universal GUT boundary conditions fail to explain the observed diphoton signal (see also~\cite{Tang:2015eko,Dutta:2016jqn}).\footnote{See~\cite{Patel:2015ulo,Ko:2016lai,Chao:2016mtn,Deppisch:2016scs,Dorsner:2016ypw,Aydemir:2016qqj,King:2016wep,Li:2016xcj} for other attempts to explain the diphoton excess within GUTs.}

We are aiming to extend the analysis~\cite{Hall:2015xds} in two ways: first we perform a full likelihood analysis on the ATLAS and CMS data in order to pin down the experimentally favored diphoton cross section. Inclusion of the 8 TeV data reduces the best fit cross section significantly and simplifies the explanation of the excess in terms of a GUT. Second we take into account sfermionic contributions to the diphoton cross section which can be significant (this was also pointed out in~\cite{Dutta:2016jqn}).

While large couplings between the 750 GeV boson and the vector sfermions can drive a strong diphoton signal, they may induce undesirable charge breaking vacua. We derive upper limits on the scalar couplings by requiring that the universe is at least metastable. For this we explicitly calculate tunneling rates along the most dangerous directions in fields space. Despite the constraints, we show that light sfermions in the loop can enhance the diphoton cross section by up to a factor of seven. This reopens the exciting possibility of fitting the diphoton excess via a single $5+\overline{5}$ or a single $10+\overline{10}$.

The considered class of supersymmetric models does not give rise to the (experimentally slightly preferred~\cite{ATLAS:13tev}) large decay width of the diphoton resonance. However, it is possible to employ the superposition of the scalar and the pseudoscalar singlet~\cite{Hall:2015xds}. Our likelihood analysis suggests that -- given a mass splitting of $\mathcal{O}(50\:\text{GeV})$ -- the corresponding double resonance fits the ATLAS and CMS data with the same quality as a broad resonance. 

Finally, we predict the signals in the $\gamma Z$, $ZZ$ and $WW$ channels which will soon come into reach at LHC-13.

\section{Signal Analysis}\label{sec:signal}

\begin{figure}[htp]
\begin{center}
\includegraphics[height=6.8cm]{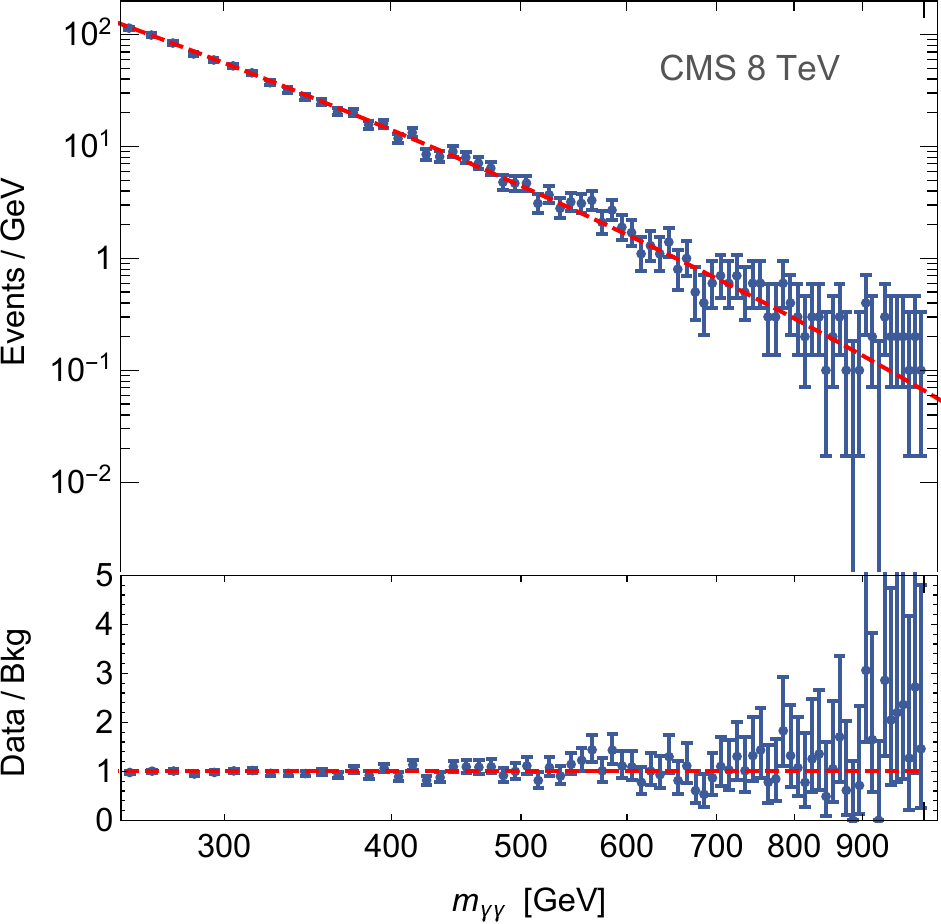}\hspace{5mm}
\includegraphics[height=6.8cm]{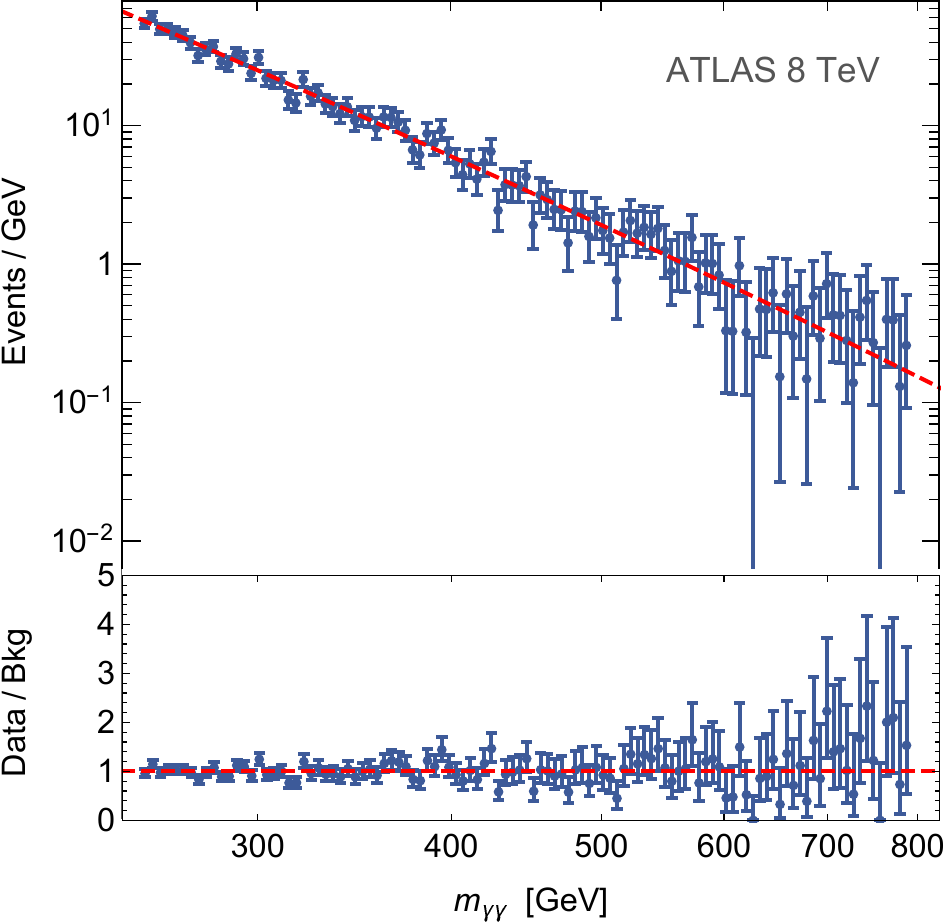}\\[2mm]
\includegraphics[height=6.8cm]{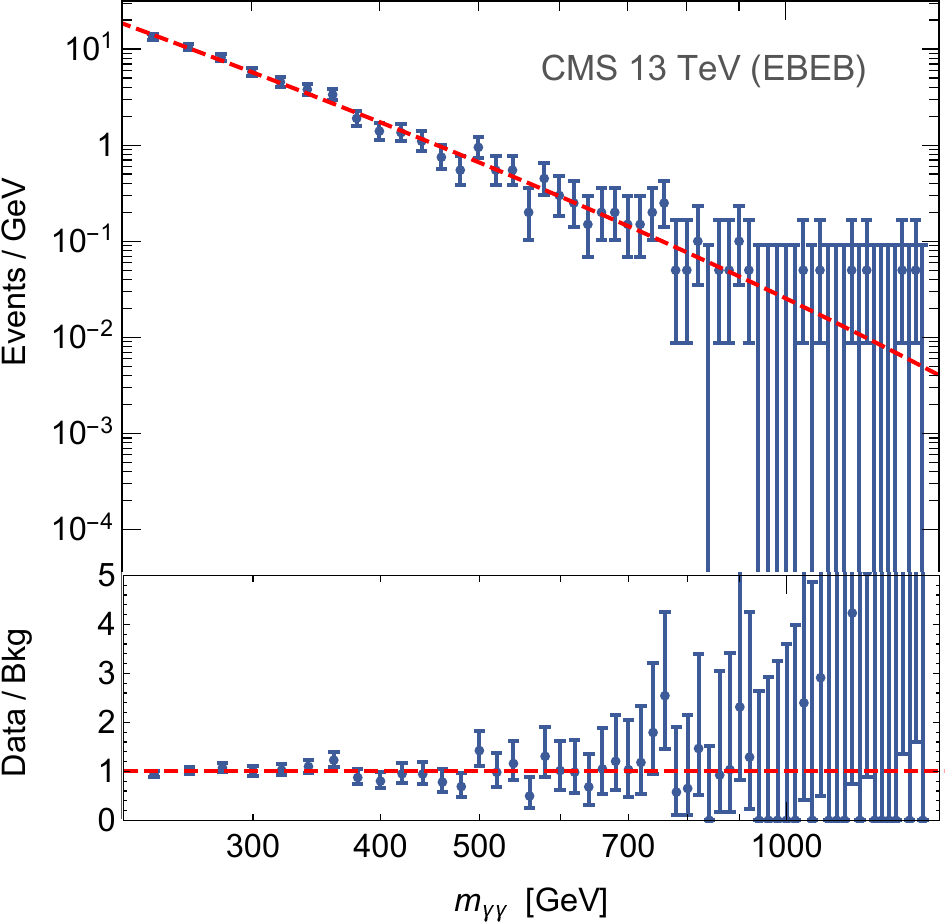}\hspace{5mm}
\includegraphics[height=6.8cm]{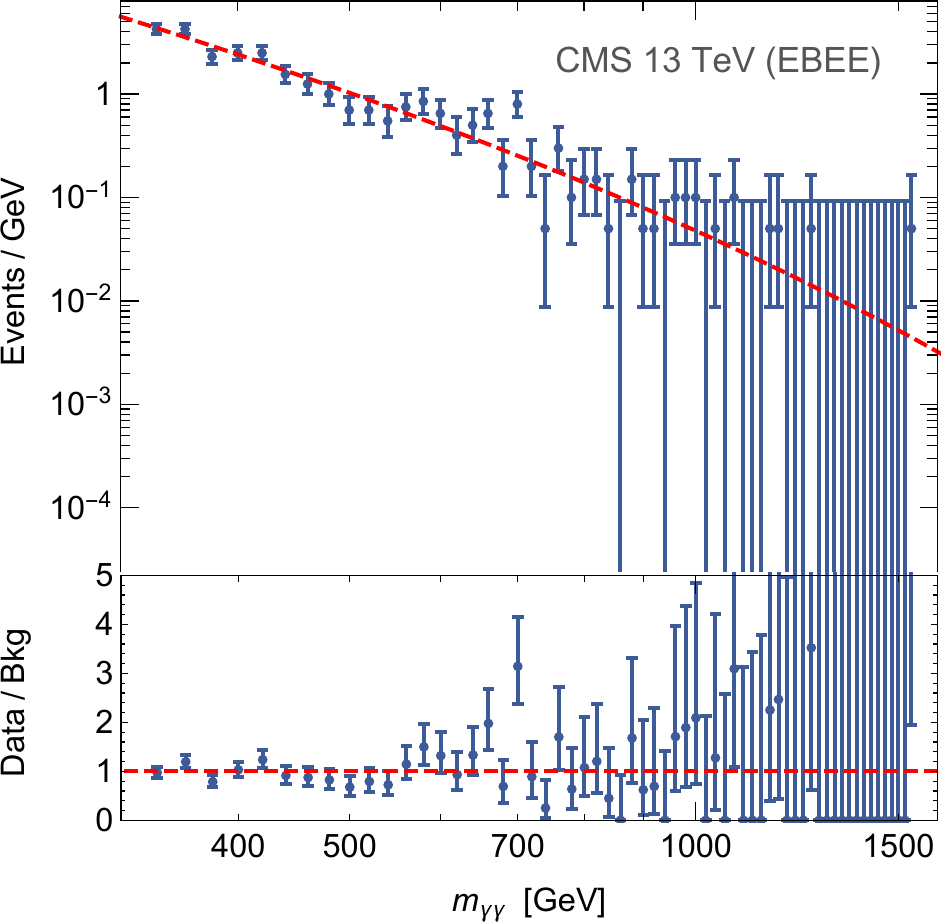}\\[2mm]
\includegraphics[height=6.8cm]{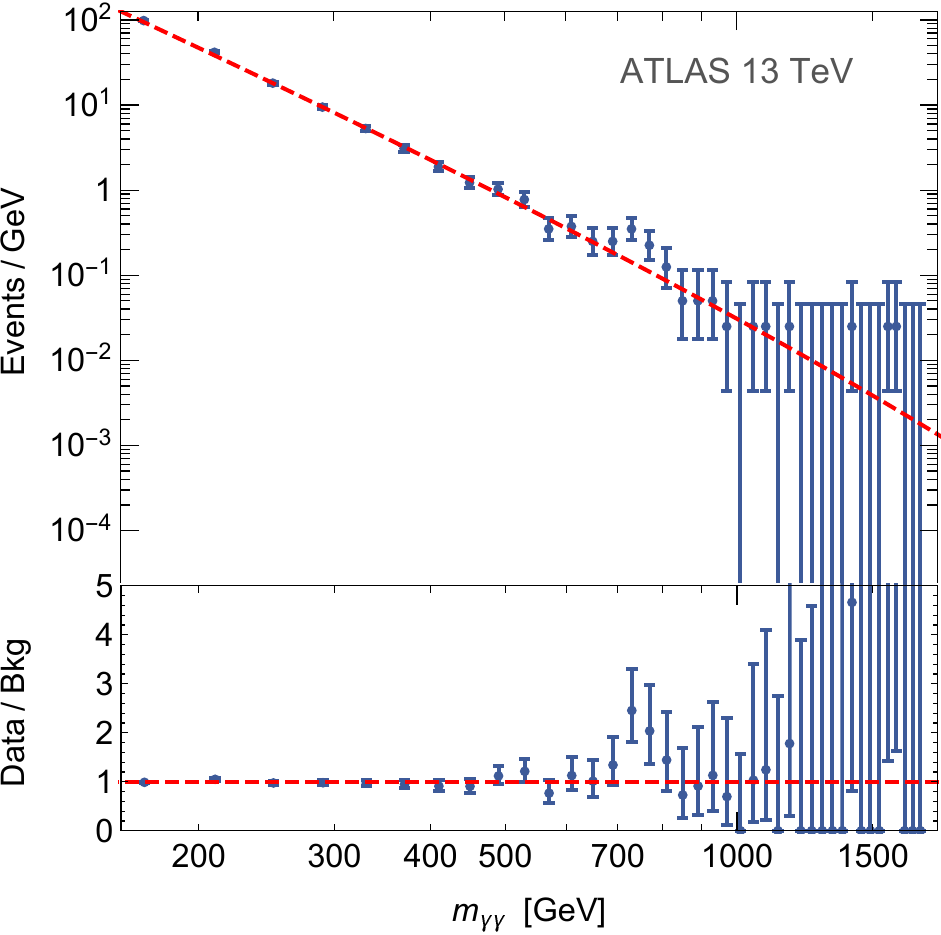}
\end{center}
\caption{Diphoton data of ATLAS and CMS and backgrounds (Bkg) employed in this work.}
\label{fig:data}
\end{figure}

In this section we perform a likelihood analysis on the diphoton spectrum. We include the ATLAS and CMS searches for diphoton resonances at 13 TeV~\cite{ATLAS:13tev,CMS:2015dxe} and 8 TeV~\cite{Aad:2014ioa,Khachatryan:2015qba}. Backgrounds are determined by fitting a smooth function to the data. As the experimental analyses were performed by different groups, the background modeling differs slightly among them. In our combined analysis, we decided to parameterize all backgrounds in a uniform way~\cite{ATLAS:13tev,Kavanagh:2016pso}
\begin{equation}
f(m_{\gamma\gamma}) = \mathcal{N} \left(1-\left(\frac{m_{\gamma\gamma}}{\sqrt{s}}\right)^{1/3}\right)^a  \left(\frac{m_{\gamma\gamma}}{\sqrt{s}}\right)^b\,,
\end{equation}
where $m_{\gamma\gamma}$ denotes the diphoton invariant mass and $\sqrt{s}$ the center-of-mass energy. For each data set the free parameters $\mathcal{N}$, $a$ and $b$ are determined by a fit to the experimental data. This is done by minimizing the Poissonian likelihood
\begin{equation}\label{eq:Lb}
L_b =\prod\limits_i \frac{\left(N^i_\text{b}\right)^{N_o^i}}{N_o^i !} e^{-N^i_\text{b}}\,,
\end{equation}
where $N_o^i$ denotes the observed number and $N_b^i$ the expected number of events (according to the background hypothesis) in the bin $i$. In order to avoid contamination of the background fit by a signal component, we exclude the bins between 700 and 800 GeV in this procedure. The diphoton data and our background fits are shown in figure~\ref{fig:data}.

For the signal, we assume a spin 0 resonance produced via gluon fusion. This implies an enhancement of the cross section at 13 TeV by a factor of 4.7 compared to 8 TeV~\cite{ATLAS:13tev,CMS:2015dxe} (our results are always given in terms of the 13 TeV cross section). The signal has to be weighted by the detector acceptance which is provided by the experimental collaborations~\cite{Aad:2014ioa,Khachatryan:2015qba,ATLASMoriond,CMSMoriond}.

Three different signal hypothesis are considered: a narrow resonance, a broad resonance and a double resonance.\footnote{A likelihood analysis on the diphoton data was also performed in~\cite{Buckley:2016mbr}. Differences compared to our analysis include the background modeling and the case of the double resonance which was not considered in~\cite{Buckley:2016mbr}. The double resonance scenario was considered in~\cite{Cao:2016cok}, however, it was only tested against the ATLAS 13 TeV data.} In the narrow resonance case, we assume that the signal width is dominated by the experimental resolutions. The latter are modeled as Gaussian with the standard deviations taken from the experimental publications. This is a slight simplification compared to the more elaborate crystal ball functions employed by the experimentalists. For the broad width resonance, we neglect the detector resolution against the intrinsic decay width which follows a Breit-Wigner distribution. Finally for the double resonance case we assume the appearance of two narrow resonances in close proximity. This scenario is e.g. realized if the new boson is a complex scalar field with its scalar and pseudoscalar components split by loop-effects. It can, in particular, be accommodated in supersymmetric models~\cite{Hall:2015xds}.

\begin{figure}[htp]
\begin{center}
\includegraphics[width=12.7cm]{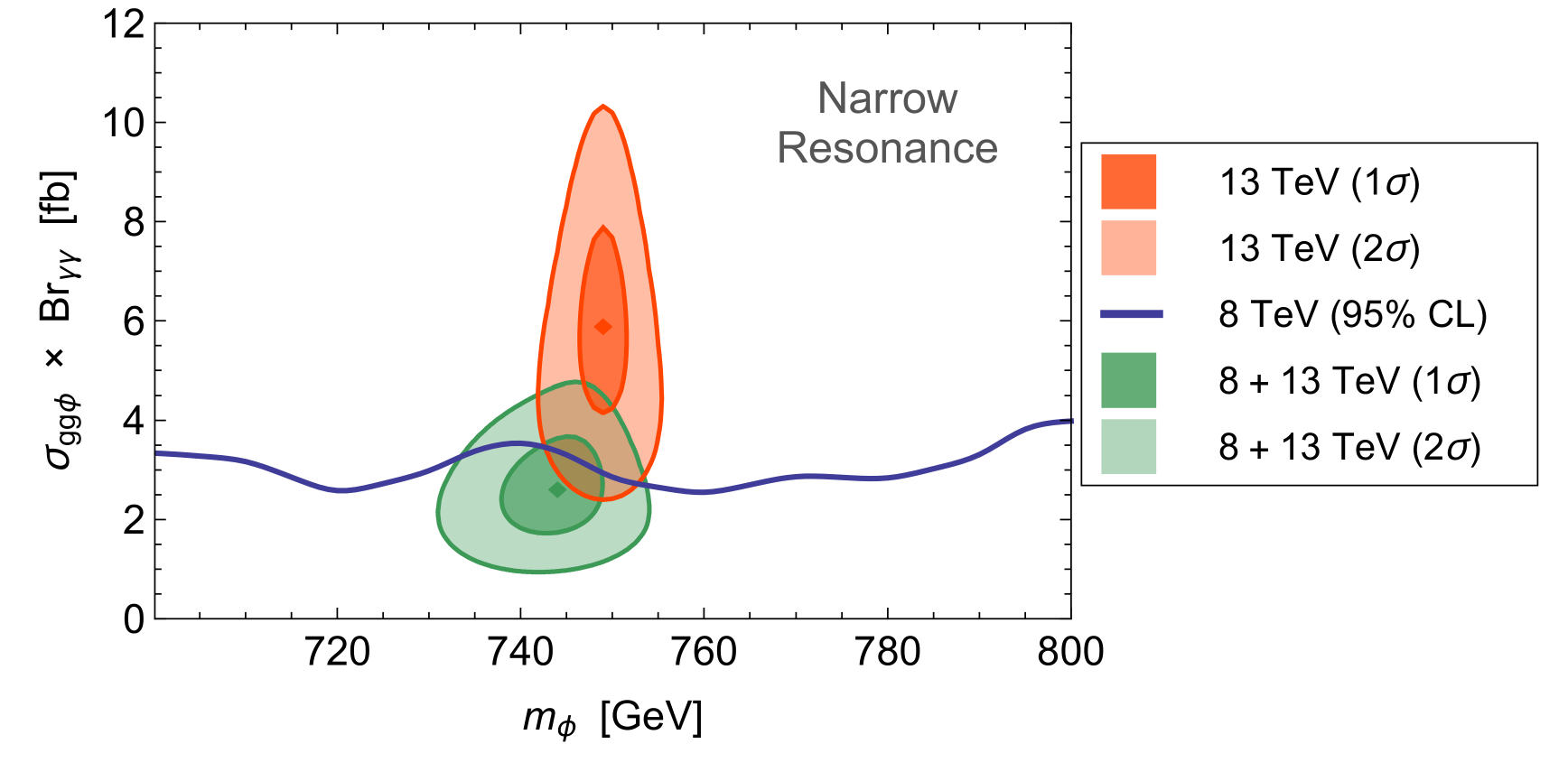}\\[3mm]
\includegraphics[width=12.7cm]{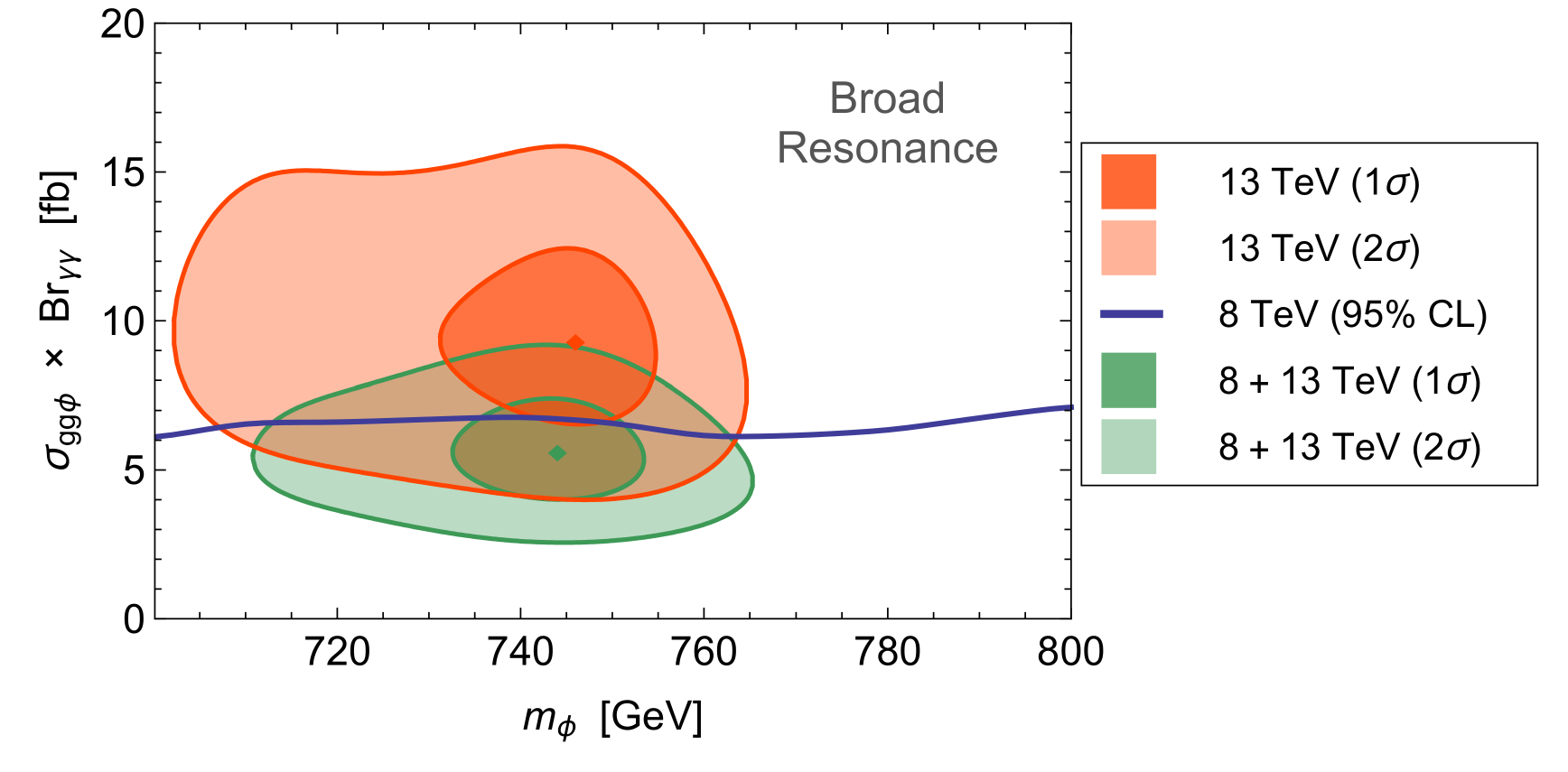}\\[3mm]
\includegraphics[width=12.7cm]{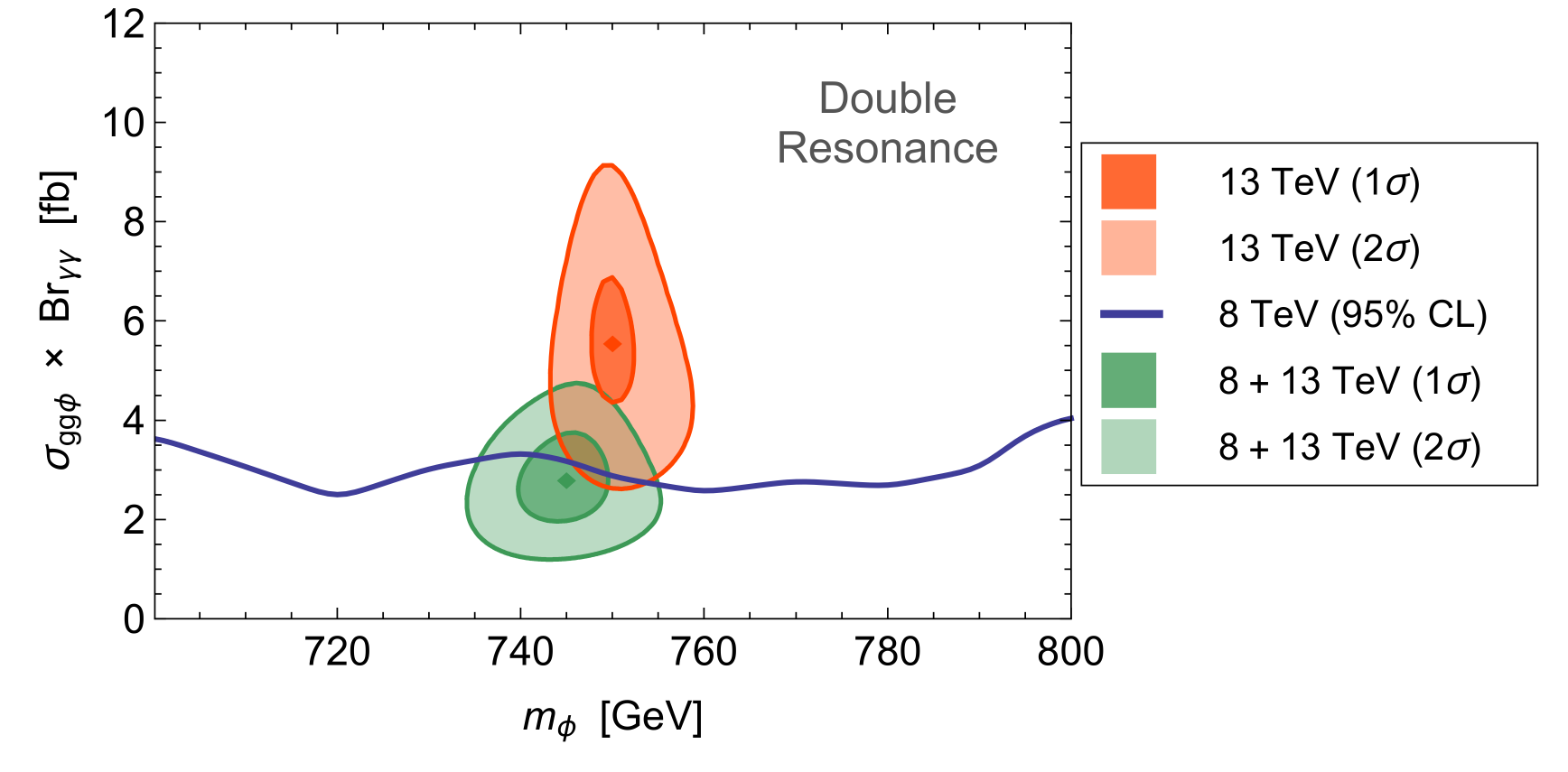}
\end{center}
\caption{Confidence regions in the plane of resonance mass and diphoton cross section for the 13 TeV data alone and the combination of 8 + 13 TeV data. Also shown are the 95\% CL upper limits from the 8 TeV data. It is assumed that the resonance is produced via gluon fusion. The cases of a narrow, broad and double resonance refer to the upper, middle and lower panel respectively. Assumptions in the broad and double resonance cases are described in the text.}
\label{fig:confidence}
\end{figure}

For each of the three scenarios, we have identified the best fit cross section for the 13 TeV data alone and for the combination of 8 and 13 TeV data. This was done by maximizing the likelihood $L_{s+b}$ of the signal plus background hypothesis which is defined in analogy to~\eqref{eq:Lb}.
The confidence intervals around the best fit point are then determined via a likelihood ratio test employing the ratio $L_{s+b}/(L_{s+b})_\text{best-fit}$. The 8 TeV data show no clear indication of a signal and are thus used to provide a $95\%$ CL upper limit on the diphoton cross section. In figure~\ref{fig:confidence} we provide the resulting confidence intervals and upper limits. For the broad width, we fixed $\Gamma_\phi = 40\:\text{GeV}$. For the double resonance, we assumed a mass splitting\footnote{The sign of the mass splitting is chosen such that the dominant resonance is heavier.} of $\Delta m = 40\:\text{GeV}$ and a relative normalization of 9:4 of the two peaks (as expected for pseudoscalar and scalar resonances originating from the same complex field). The shown mass and cross section in this case refer to the dominant resonance.

The favored cross section is considerably smaller for the narrow and double resonance compared to the broad resonance which spreads the signal among more bins. Notice also that there is some tension between the 8 TeV and the 13 TeV data, with the latter favoring significantly larger cross sections. This tension is somewhat reduced by the Moriond 2016 update which was not yet taken into account. Still, it is clear that if the diphoton signal is real, the 13 TeV data correspond to an upward fluctuation.

In table~\ref{tab:significance} we compare the overall local significance of the excess for the narrow, broad and double resonance. For the broad resonance we consider a fixed width of $\Gamma_\phi=40\:\text{GeV}$ as well as a free width $\Gamma_\phi\leq 100\:\text{GeV}$. In the double resonance case we fit the mass splitting and consider a fixed relative normalization of 9/4 as well as a free normalization. 

\begin{table}[ht]
\begin{center}
\begin{tabular}{|c|cccc|}
\hline 
 \rowcolor{light-gray}&&&&\\[-3mm]
 \rowcolor{light-gray}Resonance Type& $m_\phi$ & $\sigma_{gg\phi}\times\text{Br}_{\gamma\gamma}$ & $\Gamma_\phi$ &  Significance\\[1mm]
\hline
&&&&\\[-3mm]
Narrow & $744\:\text{GeV}$  & $2.6\:\text{fb}$ &  $-$ & $3.3\,\sigma$\\[1mm]
Broad (fixed width) & $744\:\text{GeV}$  & $5.6\:\text{fb}$ & $40\:\text{GeV}$ & $3.9\,\sigma$\\[1mm]
Broad (free width) & $745\:\text{GeV}$  & $6.9\:\text{fb}$ & $62\:\text{GeV}$  & $4.0\,\sigma$\\[1mm]
Double (fixed ratio) & $745\:\text{GeV} \;\;(705\:\text{GeV}) $  & $2.8\:\text{fb}\;\;(1.3\:\text{fb})$ & $-$ & $3.8\,\sigma$\\[1mm]
Double (free ratio) & $745\:\text{GeV} \;\;(706\:\text{GeV}) $  & $2.5\:\text{fb}\;\;(1.8\:\text{fb})$ & $-$  & $3.9\,\sigma$\\ \hline
\end{tabular}
\end{center}
\caption{Best Fit Points for a narrow, broad and double resonance and corresponding (local) significance of the excess. For the double resonance the values in brackets refer to the subdominant resonance}
\label{tab:significance}
\end{table}

The narrow width reaches a significance of $3.3\,\sigma$ which is improved to $4\,\sigma$ for the broad resonance. The double resonance yields a similar quality of fit as the broad resonance. In both cases, the gain in significance in insufficient to discard the narrow width hypothesis -- in particular as it comes at the prize of introducing one/ two new parameters. However, it is encouraging that broad and double resonance exhibit comparable significance as the latter can much easier be realized in perturbative models explaining the diphoton excess.

\section{Diphotons in Supersymmetry}

We will now turn to the explanation of the diphoton excess in terms of supersymmetric models. The minimal supersymmetric standard model (MSSM) itself fails to provide a candidate for the 750 GeV boson (see however~\cite{Bharucha:2016jyr}).\footnote{In the NMSSM, the diphoton signal can be mimicked by the decay of two ultra light pseudoscalars into collimated photons~\cite{Ellwanger:2016qax,Domingo:2016unq,Badziak:2016cfd}.} Therefore, we shall consider a simple extension of the MSSM by a singlet $S$ and vector-like chiral superfields $X_i$~\cite{Hall:2015xds}
\begin{equation}
W = W_\text{MSSM} + \frac{\mu_s}{2} S^2 + \mu_i\,\overline{X}_i X_i + \lambda_i\, S\, \overline{X}_i X_i\,.
\end{equation}
The singlet is taken to be even, the vector-like fields odd under R-parity. Provided the $X_i$ come in complete SU(5) multiplets, the attractive feature of gauge coupling unification is preserved. However, any new multiplet increases the unified gauge coupling. Requiring unification in the perturbative regime one may introduce up to three copies of $5+\overline{5}$ 
or one $10+\overline{10}$. The 5-plet contains a down-type quark and a lepton doublet $5 = (D,\overline{L})$, while the 10-plets contains a quark doublet, an up-type quark and a lepton singlet $10 = (Q,\overline{U},\overline{E})$. In order to avoid stable vector-like particles, the superpotential should be extended by Yukawa couplings of the $X_i$ to the MSSM fields. We assume these couplings to be present, but sufficiently suppressed not to affect our analysis. 

The complex scalar singlet decomposes into a CP even state $h_s$ and a CP odd state $a_s$. The tree-level masses read
\begin{equation}
m_{h_s}^2= m_s^2 + \mu_s^2 + B\mu_s\,, \qquad m_{a_s}^2= m_s^2 + \mu_s^2 - B\mu_s\,,
\end{equation} 
where $m_s$ and $B\mu_s$ denote the singlet soft mass and bilinear soft breaking term. The $750\:\text{GeV}$ resonance can be identified either with $h_s$, $a_s$ or the superposition of both. The last possibility can be realized if the mass splitting between scalar and pseudoscalar $|m_{h_s}-m_{a_s}|\lesssim 50\:\text{GeV}$. This translates into $\sqrt{|B\mu_s|} \lesssim 200\:\text{GeV}$. If $B\mu_s$ is suppressed one additionally has to take into account loop effects which increase or decrease the splitting by a few tens of GeV.\footnote{Loop effects also induce a small vev of $h_s$ through the tadpole diagram with vector-like states in the loop.} Depending on whether the mass splitting exceeds the detector resolution, scalar and pseudoscalar are either observed as a single or as a double resonance.

The production and decay of the resonance proceeds via loops of vector-like fermions or sfermions. The vector quarks and leptons form Dirac fermions of mass $\mu_i$ with their coupling to $h_s$ and $a_s$ given by $\lambda_i$. Scalar masses also receive contributions from supersymmetry breaking. The sfermion mass matrix in the basis $(\widetilde{X},\widetilde{\overline{X}}^*)$ with $X=Q,\,U,\,D,\,L,\,E$ reads
\begin{equation}\label{eq:sfermionmassmatrix}
M_{\widetilde{X}} = \begin{pmatrix}
\mu_X^2 +m_X^2 & B \mu_X \\
B \mu_X & \mu_X^2 +m_{\overline{X}}^2 
\end{pmatrix}
\,,
\end{equation}
with the supersymmetric mass $\mu_X$, the soft masses $m_X$, $m_{\overline{X}}$ and the bilinear soft breaking term $B \mu_X$. We neglected small electroweak corrections to keep the SU(2) doublets degenerate. The mass eigenstates will be denoted by $\widetilde{X}_{1,2}$, where the index $1$ indicates the lighter state. They are obtained from $(\widetilde{X},\widetilde{\overline{X}}^*)$ by a rotation in field space with the mixing angle $\theta_X$.

In the presence of CP conservation, sfermions only couple to the CP even scalar $h_s$. The Lagrangian contains trilinear couplings $\mathcal{L} \supset -T_{X_i} |\widetilde{X}_i|^2 h_s $ with\footnote{Trilinear couplings $\widetilde{X}_1  \widetilde{X}_2^* h_s$ are also present but not relevant for our analysis.}
\begin{equation}\label{eq:trilinear}
T_{X_1} = \lambda_X \frac{2 \mu_X + (A_X+\mu_s) \sin(2\theta_X)}{\sqrt{2}}\,,\qquad
T_{X_2} = \lambda_X \frac{2 \mu_X - (A_X+\mu_s) \sin(2\theta_X)}{\sqrt{2}}\,.
\end{equation}
The term in brackets originates from supersymmetry breaking and is maximized for a large mixing angle $\theta_X \sim 45^\circ$. Notice that it carries opposite sign for the lighter and heavier mass eigenstate. 

Let us now turn to the production of the singlet scalar or pseudoscalar $\phi=h_s,\,a_s$. Assuming gluon fusion, the production cross section at next-to-next-to-leading order (NNLO) is obtained from the leading order (LO) cross section by multiplying with the relevant $k$-factor. The latter is estimated by taking the ratio of NNLO and LO cross sections of the standard model Higgs $ \sigma_{ggh}^\text{NNLO}/ \sigma_{ggh}^\text{LO}$ (with the Higgs mass set to $m_h=m_\phi$). This leads to the expression
\begin{equation}
\sigma_{gg\phi}^\text{NNLO} \simeq \frac{\sigma_{ggh}^\text{NNLO}}{\sigma_{ggh}^\text{LO}}\times \sigma_{gg\phi}^\text{LO} \simeq \frac{\Gamma_{\phi\rightarrow gg}^\text{LO}}{\Gamma_{h\rightarrow gg}^\text{LO}}
\times \sigma_{ggh}^\text{NNLO}\,,
\end{equation}
where we made use of the correspondence between gluonic production and gluonic decay of the resonance. The NNLO cross section of a heavy standard model Higgs is determined with the tool SusHi (version 1.5.0)~\cite{Harlander:2012pb} employing the MSTW2008 PDFs~\cite{Martin:2009iq}.

The (leading order) decay rate of the resonance into two-photon and two-gluon final states is given as~\cite{Spira:1995rr}
\begin{subequations}\label{eq:diphogludecay}
\begin{align}
\Gamma_{\phi\rightarrow\gamma\gamma}^\text{LO}&= \frac{\alpha^2 m_\phi^3}{256\pi^3}\left|
\sum\limits_f \frac{N_c^f \,\lambda_f \,Q_f^2}{\sqrt{2} \,m_f}\;A_f\!\left(\frac{m_\phi^2}{4m_f^2}\right) +
\sum\limits_s \frac{N_c^s \,T_s\, Q_s^2}{2 \,m_s^2}\;A_s\!\left(\frac{m_\phi^2}{4m_s^2}\right)
\right|^2\,,\\
\Gamma_{\phi\rightarrow gg}^\text{LO}&= \frac{\alpha_s^2 m_\phi^3}{128\pi^3}  \,\left|
\sum\limits_{f\,\in\; \text{colored}} \frac{\lambda_f}{\sqrt{2} \,m_f}\;A_f\!\left(\frac{m_\phi^2}{4m_f^2}\right) +
\sum\limits_{s\,\in\; \text{colored}} \frac{T_s}{2 \,m_s^2}\;A_s\!\left(\frac{m_\phi^2}{4m_s^2}\right)
\right|^2\,,
\end{align}
\end{subequations}
with the indices $f$ and $s$ running over the vector-like fermions and scalars (only the colored ones in the case of $gg$). Electric charges are denoted by $Q$ and the color factors are $N_c=3$ for quarks and $N_c=1$ for leptons. We set $\alpha=0.0078$ and $\alpha_s= 0.095$~\cite{Chatrchyan:2013txa}. The loop functions are defined as
\begin{subequations}
\begin{align}
A_f(\tau) &= \begin{cases}
2\tau^{-2} \,\left(\tau + (\tau - 1)\arcsin^2(\sqrt{\tau})\right)\phantom{0}&(\phi=h_s)\,,\\
2\tau^{-1} \arcsin^2(\sqrt{\tau})\;\;&(\phi=a_s)\,,
\end{cases}\\
A_s(\tau) &= \begin{cases}
\tau^{-2} \,\left(\arcsin^2(\sqrt{\tau})-\tau\right) \hspace{1.7cm}\; &(\phi=h_s)\,,\\
0&(\phi=a_s)\,,\\
\end{cases}
\end{align}
\end{subequations}
where we assumed $m_{f,s} > m_\phi /2 $ such that two-body decays of the resonance into vector (s)fermions are forbidden.

For the decay into gluons, we take into account NLO and NNLO corrections~\cite{Chetyrkin:1997iv} which increase the width by a factor $\sim$1.7. The remaining diboson decay rates $\Gamma_{\phi\rightarrow xx}$ are obtained from $\Gamma_{\phi\rightarrow\gamma\gamma}$ by replacing
\begin{align}
\quad Q_{f,s}^2 \quad\longrightarrow\quad 
\begin{cases}
\sqrt{2}\, (-I^3_{f,s}\cot\theta_W + Y_{f,s}\tan\theta_W)\, Q_f\,,\qquad &xx=\gamma Z\,,\\[2mm]
\left(-I^3_{f,s}\cot\theta_W + Y_{f,s}\tan\theta_W\right)^2\,, \qquad &xx=ZZ\,,\\[2mm]
\sqrt{2}\, \left(\frac{I^3_{f,s}}{\sin\theta_W }\right)^2\,,\qquad &xx=WW\,.\\
\end{cases}
\end{align}
Hypercharge and third component of isospin are denoted by $Y$ and $I^3$ respectively. Given the (unrealistic) case that $\Gamma_\phi$  is dominated by one particular vector-like fermion $X =Q,\,U,\,D,\,L,\,E$ or its superpartner in the loop, the diboson decay pattern would be completely fixed (see table~\ref{tab:simpledecay}). In reality, we expect several vector-like states to contribute to $\Gamma_\phi$. The decay pattern then depends on their relative couplings and masses. We will come back to this point in section~\ref{sec:predictions}. Note also that the CP even singlet can decay into $WW,\,ZZ$ at tree-level via mixing with the standard model Higgs. In our case, the mixing is highly suppressed as it is only induced at loop-order. However, if there exists a coupling $\lambda S H_u H_d$ in the superpotential, tree-level decays into $WW,\,ZZ$ easily become significant.

\begin{table}[ht]
\begin{center}
\begin{tabular}{|c|cccc|}
\hline 
\rowcolor{light-gray}&&&&\\[-3mm]
\rowcolor{light-gray}Dominant Loop& $\Gamma_{\phi\rightarrow Z\gamma}/\Gamma_{\phi\rightarrow\gamma\gamma}$ & $\Gamma_{\phi\rightarrow ZZ}/\Gamma_{\phi\rightarrow\gamma\gamma}$ &$\Gamma_{\phi\rightarrow WW}/\Gamma_{\phi\rightarrow\gamma\gamma}$& $\Gamma_{\phi\rightarrow gg}/\Gamma_{\phi\rightarrow\gamma\gamma}$\\[1mm]
\hline
&&&&\\[-3mm]
$D$ & $0.6$  & $0.1$ & $-$ & $3933$ \\[1mm]
$L$ & $0.8$  & $3.3$ & $9.3$ & $-$ \\[1mm]
$Q$ & $5.0$  & $9.1$ & $30$ & $629$ \\[1mm]
$U$  & $0.6$  & $0.1$ & $-$ & $246$ \\[1mm]
$E$  & $0.6$  & $0.1$ & $-$ & $-$ \\ \hline
\end{tabular}
\end{center}
\caption{Relative diboson decay widths of the $750\:\text{GeV}$ resonance under the assumption that the width is completely dominated by one vector-like state ($D$, $L$, $Q$, $U$ or $E$) in the loop.}
\label{tab:simpledecay}
\end{table}

The diphoton cross section is defined as $\sigma_{gg\phi}\times \text{Br}_{\gamma\gamma}$ with the branching ratio
\begin{equation}
\text{Br}_{\gamma\gamma} = \frac{\Gamma_{\phi\rightarrow\gamma\gamma}}{\Gamma_{\phi\rightarrow\gamma\gamma}+\Gamma_{\phi\rightarrow Z\gamma}+\Gamma_{\phi\rightarrow ZZ}+\Gamma_{\phi\rightarrow WW}+\Gamma_{\phi\rightarrow gg}}\;\;.
\end{equation}
While the denominator is typically dominated by $\Gamma_{\phi\rightarrow gg}$, other diboson channels may become important if the (s)leptons are considerably lighter than the (s)quarks.

\section{Maximal Diphoton Cross Section}

We will now determine the maximal diphoton cross section in the MSSM extended by up to three $5+\overline{5}$ or one $10+\overline{10}$. Assuming universal boundary conditions at the GUT scale, the ratio of fermion masses and couplings at the low scale is then predictable through the renormalization group equations (RGEs). In order to explain the diphoton signal, large couplings $\lambda_i$ of the singlet to the vector-like fermions would be desirable. Unfortunately, these couplings are tightly constrained by requiring perturbativity up to the GUT scale. Too large $\lambda_i$ at the low scale result in Landau poles in the renormalization group running.\footnote{See~\cite{Bae:2016xni} for model-independent constraints.} The maximal couplings consistent with grand unification are obtained from the two-loop RGEs by setting $\lambda_i=\sqrt{4\pi}$ as initial condition at the GUT scale $M_\text{GUT}= 2 \cdot 10^{16}\:\text{GeV}$.\footnote{We generated the two-loop RGEs with the tool SARAH (version 4.8.1)~\cite{Staub:2008uz,Staub:2013tta}.} They are shown in table~\ref{tab:yukawas}.  The numbers can be compared with those from the one-loop analysis~\cite{Hall:2015xds}. Deviations reside at the level of $\leq 10\%$ and can be traced back to two-loop effects.

\begin{table}[ht]
\begin{center}
\begin{tabular}{|c|ccccc|ccc|}
\hline 
\rowcolor{light-gray}&&&&&&&&\\[-3mm]
\rowcolor{light-gray}Vector-like Particles & $\lambda_D$ & $\lambda_L$ &  $\lambda_Q$ &$\lambda_U$ &$\lambda_E$ & $\mu_D/\mu_L$ & $\mu_Q/\mu_E$ & $\mu_U/\mu_E$\\[1mm]
\hline
&&&&&&&&\\[-3mm]
$(5+\overline{5})_1$ & $0.97$ & $0.63$ & $-$ & $-$ & $-$ & $1.55$ & $-$ & $-$ \\[1mm]
$(5+\overline{5})_2$ & $0.79$ & $0.45$ & $-$ & $-$ & $-$ & $1.75$ & $-$ & $-$ \\[1mm]
$(5+\overline{5})_3$ & $0.71$ & $0.34$ & $-$ & $-$ & $-$ & $2.09$ & $-$ & $-$ \\[1mm]
$10+\overline{10}$& $-$ & $-$ & $0.88$ & $0.71$ & $0.26$ & $-$ & $3.32$ & $2.70$ \\ \hline
\end{tabular}
\end{center}
\caption{Maximal couplings of the vector-like fermions consistent with perturbativity up to the GUT scale. Fermion mass ratios are also given.}
\label{tab:yukawas}
\end{table}

Due to an infrared fixed-point in the RGEs of the $\lambda_i$, couplings at the low scale are rather insensitive to the high scale input. Any choice of $\lambda_i \gtrsim  1$ at the GUT scale results in fermion couplings very close to those of table~\ref{tab:yukawas}. The table also contains the fermion mass ratios in the fixed-point scenario which are again determined by the two-loop RGEs.

Let us for the moment assume that the sfermions receive large soft masses such that they decouple. In this case only fermion-loops contribute to the diphoton signal. Taking couplings and mass ratios from table~\ref{tab:yukawas}, there remains one free parameter which can be chosen to be the vector lepton mass ($\mu_L$ or $\mu_E$). Cross sections of the scalar and pseudoscalar singlet only differ by the loop function $A_f$ which approaches the constants 4/3 (scalar) and 2 (pseudoscalar) in the heavy fermion limit (fermions heavier than $500\:\text{GeV}$). This translates into the ratio $\sigma_{gg h_s} /\sigma_{gg a_s} = 4 /9$. The signal strength is maximized if scalar and pseudoscalar both contribute to the diphoton excess.

\begin{figure}[t]
\begin{center}
\includegraphics[width=11cm]{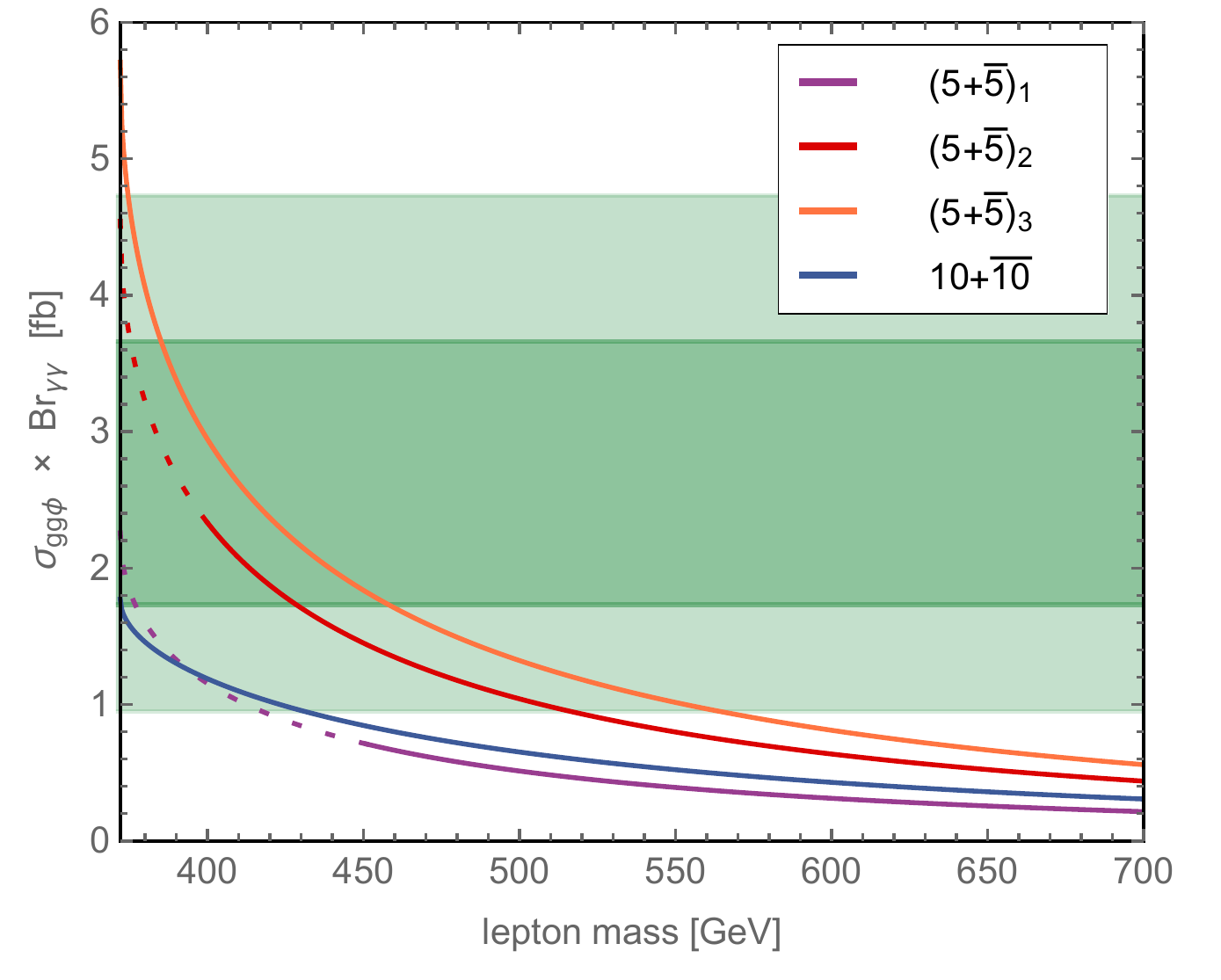}
\end{center}
\caption{Maximal diphoton cross section induced by vector-like fermion loops if perturbativity up to the GUT scale is imposed. Vector-like fields are assumed to be in complete SU(5) multiplets (up to three $5+\overline{5}$ or one $10+\overline{10}$) with universal boundary conditions at the GUT scale. The vector sfermions are taken to be decoupled via large soft masses. For the case of one or two $5+\overline{5}$, the dashed line indicates that $\mu_D< 700\:\text{GeV}$ which is considered to be in tension with direct quark searches at the LHC. The experimentally preferred diphoton cross section from our likelihood analysis ($1$ and $2\,\sigma$ band) is shown in green.}
\label{fig:fermion}
\end{figure}

In figure~\ref{fig:fermion} we depict the diphoton cross section as a function of the lepton mass for a given vector-like field content. We set $m_{h_s}=m_{a_s}=744\:\text{GeV}$. Lines in the figure are dotted if the lightest vector quark mass falls below $700\:\text{GeV}$ which we take as the lower limit from direct searches for heavy quarks at the LHC. Depending on the decay modes of the vector fermions the actual limit varies, but is expected to lie in this ballpark (see e.g.\ \cite{Aad:2015mba,Aad:2015kqa,Aad:2015tba}). We also show the experimentally preferred cross section from our likelihood analysis in the narrow width case. 

Encouragingly, we observe that two $5+\overline{5}$ or one $10+\overline{10}$ can already fit the excess at the $1\,\sigma$ level -- even before we included sfermion loops. We are thus more optimistic than~\cite{Hall:2015xds} despite the fact that our loop calculations are in good agreement\footnote{We obtain cross sections $\sim 20\%$ smaller than in~\cite{Hall:2015xds}. This corresponds roughly to the reduction of the diphoton cross section by NNLO corrections to the two-gluon decay of the resonance.}. The key difference is that~\cite{Hall:2015xds} assumed a larger experimentally preferred cross section $\sigma_{gg\phi}\times \text{Br}_{\gamma\gamma}=4-8\:\text{fb}^{-1}$ which corresponds approximately to the region favored by LHC-13 (see our figure~\ref{fig:confidence}). However, as pointed out in section~\ref{sec:signal}, consistency of the 8 and 13 TeV data requires an upward fluctuation at 13 TeV. The combined data favor the smaller cross sections employed in our analysis.

Let us now discuss the sfermionic contribution to the diphoton cross section. We recall that for CP conservation, only the cross section of the CP even scalar is affected by sfermion loops. The RGE running of scalar masses and couplings depends on a large set of soft terms. Even with universal boundary conditions at the GUT scale, low scale values cannot simply be inferred. While this reduces the predictive power, it introduces some freedom to considerably enhance the diphoton cross section of the 750 GeV resonance. 

The opposite of the decoupling limit considered so far is the `supersymmetric limit', in which soft terms of the vector sfermions are suppressed against their supersymmetric masses $\mu_X$. This would lead to an increase of the cross section  $\sigma_{gg h_s}$ by a factor of $(3/2)^2$ compared to the case of decoupled sfermions. In the supersymmetric limit $\sigma_{gg h_s}$ approaches $\sigma_{gg a_s}$. If the diphoton signal originates from the superposition of scalar and pseudoscalar, sfermion loops enhance the total cross section by a factor of 18/13.

Even more interesting is the scenario, in which the sfermions carry sizeable trilinear couplings. These couplings receive contributions from supersymmetry breaking (cf.\ \eqref{eq:trilinear}) and are not directly constrained by Landau poles. For realistic high scale boundaries, the diagonal entries of the sfermion mass matrices~\eqref{eq:sfermionmassmatrix} are nearly degenerate and sizeable off-diagonal entries exist. This leads to maximal mixing ($\theta_X=45^\circ$) of the sfermion mass eigenstates $\tilde{X}_{1,2}$ such that the trilinear couplings become
\begin{equation}
T_{X_{1,2}} = \lambda_X \frac{2 \mu_X \pm (A_X+\mu_s)}{\sqrt{2}}\,.
\end{equation}
Through the term in brackets, trilinear couplings can be strongly enhanced. In particular, the sfermions may couple more strongly to the singlet than their fermionic superpartners. In this case, $T_{X_1}$ and $T_{X_2}$ carry opposite sign such that $\widetilde{X}_1$ and $\widetilde{X}_2$ interfere destructively in the sfermion loops. The diphoton cross section is hence maximized if $\widetilde{X}_2$ decouples and only $\widetilde{X}_1$ contributes.

\begin{figure}[t]
\begin{center}
\includegraphics[width=12cm]{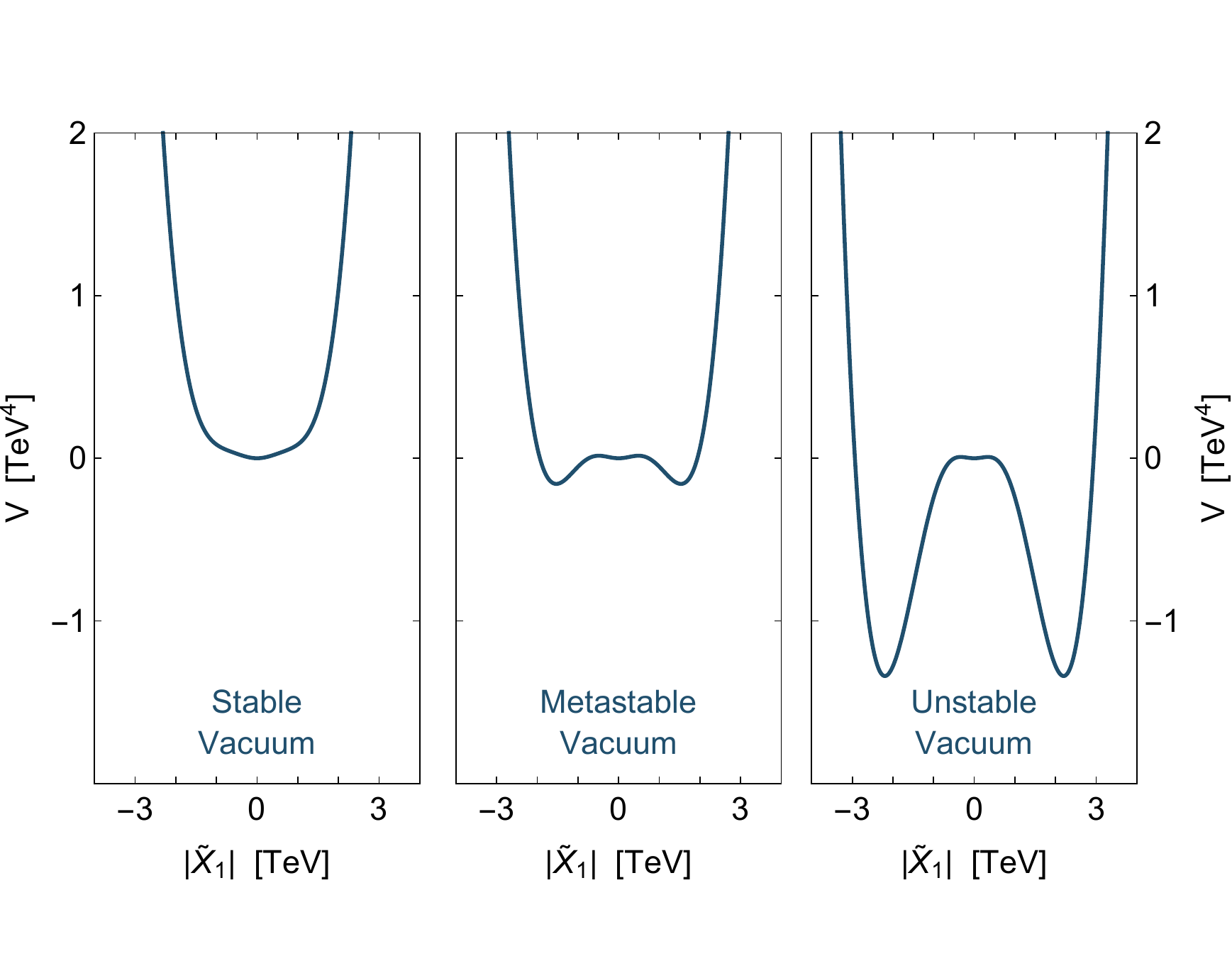}
\end{center}
\vspace{-7mm}
\caption{Scalar potential potential along the the charge breaking direction $|\widetilde{X}_1|$ for increasing scalar coupling $T_{X_1}$ (from left). The panels on the left, in the middle and on the right refer to a stable, metastable and unstable electroweak vacuum (at $\widetilde{X}_1=0$) respectively.}
\label{fig:vacua}
\end{figure}

We now want to determine the maximal coupling $T_{X_1}$. The scalar potential of the fields $\widetilde{X}_1$ and $h_s$ reads
\begin{equation}\label{eq:dangerouspotential}
  V = m_{\widetilde{X}_1}^2\: \left|\widetilde{X}_1\right|^2 + \frac{m_{h_s}^2}{2}  \:h_s^2 + T_{X_1}  \:\left|\widetilde{X}_1\right|^2 h_s + \frac{\lambda_X^2}{4} \: \left|\widetilde{X}_1\right|^4 + \frac{\lambda_X^2}{2}  \:\left|\widetilde{X}_1\right|^2 h_s^2\,,
\end{equation}
where we assumed $\theta_X=45^\circ$ resulting in a D-flat direction. This potential form holds for all $X=Q,\,U,\,D,\,L,\,E$. In case $|T_{X_1}|\gg m_{h_s},m_{\widetilde{X}_1}$ there appears an unwanted charge breaking vacuum at non-trivial field values. This can be tolerated even if the charge breaking vacuum is deeper than the standard electroweak vacuum. The universe at present does not necessarily occupy the lowest possible energy state. During the epoch of cosmic inflation, all scalar fields received inflaton-induced soft masses. Hence, the charge conserving minimum may have been preferred at early times (see e.g.\ \cite{Ellis:2008mc}). We will assume that the universe ended up in the standard electroweak vacuum even in the presence of a deeper charge breaking minimum. If such a deeper minimum exists, the universe is, however, unstable against quantum tunneling. In order to arrive at a viable model we require that the tunneling time is larger than the age of the universe.\footnote{Analog considerations were used to constrain the couplings of non-supersymmetric scalars in~\cite{Salvio:2016hnf}. In the MSSM trilinear couplings of lights stops~\cite{Kusenko:1996jn} and light staus~\cite{Ratz:2008qh} are limited by vacuum stability.} For the single field case, the tunneling rate per volume $\Gamma_\text{tunnel}$ into a deeper minimum of the potential is given as~\cite{Coleman:1977py,Callan:1977pt}
\begin{equation}
\Gamma_\text{tunnel} = A e^{-S_B}\,.
\end{equation}
where $S_B$ denotes the action of the `bounce', a classical solution to the Euler-Lagrange equation obeying certain boundary conditions~\cite{Coleman:1977py,Callan:1977pt}. Taking the prefactor $A=\mathcal{O}\left(\text{TeV}^4\right)$, a tunneling time larger than the age of the universe requires $S_B > 400$.

For the multi-field potential~\eqref{eq:dangerouspotential}, we first identify the most dangerous path in field space connecting the charge conserving and charge breaking minima. By this, we mean the path which minimizes the barrier separating the two. It is defined by
\begin{equation}\label{eq:dangerouspath}
h_s = -\frac{T_{X_1} \left|\widetilde{X}_1\right|^2}{m_{h_s}^2+\lambda_X^2 \left|\widetilde{X}_1\right|^2}\,.
\end{equation}
We eliminate $h_s$ through this equation and redefine $|\widetilde{X}_1|$ such that it is canonically normalized along this path. Finally, we determine $S_B$ by solving the differential equation of the bounce~\cite{Coleman:1977py,Callan:1977pt}. In figure~\ref{fig:vacua} we depict the scalar potential along the direction $|\widetilde{X}_1|$ with $h_s$ fixed by~\eqref{eq:dangerouspath}. The three examples refer to a stable, metastable and unstable electroweak vacuum respectively. Metastable means that the tunneling time exceeds the age of the universe. Imposing at least metastability ($S_B > 400$) results in an upper limit on the trilinear coupling $T_{X_1}$ which depends on the masses $m_{h_s},\:m_{\widetilde{X}_1}$ and the Yukawa coupling $\lambda_X$. Taking $m_{h_s}= 744 \:\text{GeV}$ and $\lambda_X$ from table~\ref{tab:yukawas} we arrive at the maximal trilinear couplings of table~\ref{tab:trilinears} for $X=Q,\,U,\,D,\,L,\,E$.

\begin{table}[ht]
\begin{center}
\begin{tabular}{|c|ccccc|}
\hline 
\rowcolor{light-gray}&&&&&\\[-3mm]
\rowcolor{light-gray}Vector-like Particles & $T_{D_1}\;[\text{TeV}]$ & $T_{L_1}\;[\text{TeV}]$ &  $T_{Q_1}\;[\text{TeV}]$ &$T_{U_1}\;[\text{TeV}]$ &$T_{E_1}\;[\text{TeV}]$ \\[1mm]
\hline
&&&&&\\[-3mm]
$(5+\overline{5})_1$ & $1.83 \:\:\:(2.34)$ & $0.95 \:\:\:(1.11)$ & $-$ & $-$ & $-$  \\[1mm]
$(5+\overline{5})_2$ & $1.57 \:\:\:(2.02)$ & $0.80 \:\:\:(0.95)$ & $-$ & $-$ & $-$  \\[1mm]
$(5+\overline{5})_3$ & $1.47 \:\:\:(1.89)$ & $0.73 \:\:\:(0.88)$ & $-$ & $-$ & $-$  \\[1mm]
$10+\overline{10}$&  $-$ & $-$ & $1.70 \:\:\:(2.19)$ & $1.47 \:\:\:(1.89)$ & $0.71 \:\:\:(0.82)$ \\ \hline
\end{tabular}
\end{center}
\caption{Maximal couplings between vector-like sfermions and CP even singlet. The maximal couplings are sensitive to the mass of the considered sfermion. For the values above (values in brackets), we set slepton masses to $380\:\text{GeV}$ ($500\:\text{GeV}$) and squark masses to $700\:\text{GeV}$ ($1000\:\text{GeV}$). The maximal coupling at a different mass can be obtained by linear extrapolation.}
\label{tab:trilinears}
\end{table}

Finally, we determine the maximal diphoton cross section including sfermion loops. The latter is obtained for maximal couplings $\lambda_X,\:T_{X_1}$ and $\mu_X,\:m_{\widetilde{X}_1}$ as light as possible. Consistency with direct searches at the LHC puts a lower limit on the masses of colored particles which depends on their decay modes. As we did not specify the couplings of the vector-like states to the MSSM fields, we cannot identify a clear-cut constraint and make the somewhat simplistic assumption that all vector (s)quarks should be heavier than $700\:\text{GeV}$ (see e.g.\ \cite{Aad:2015mba,Aad:2015kqa,Aad:2015tba}). (S)leptons can be considerably lighter, but should not be accessible as two-body final states (otherwise $\Gamma_{\phi\rightarrow\gamma\gamma}$ would be suppressed). In order to avoid fine-tuning, we  require them to be heavier than $380\:\text{GeV}$. Notice that quarks and leptons cannot both saturate the lower limit as their mass ratio is fixed (see table~\ref{tab:yukawas}). If we consider e.g.\ the vector-like 10-plet of SU(5) and impose $\mu_E > 380\:\text{GeV}$, it follows that $\mu_U > 1.03\:\text{TeV}$.

\begin{figure}[htp]
\begin{center}
\includegraphics[width=14cm]{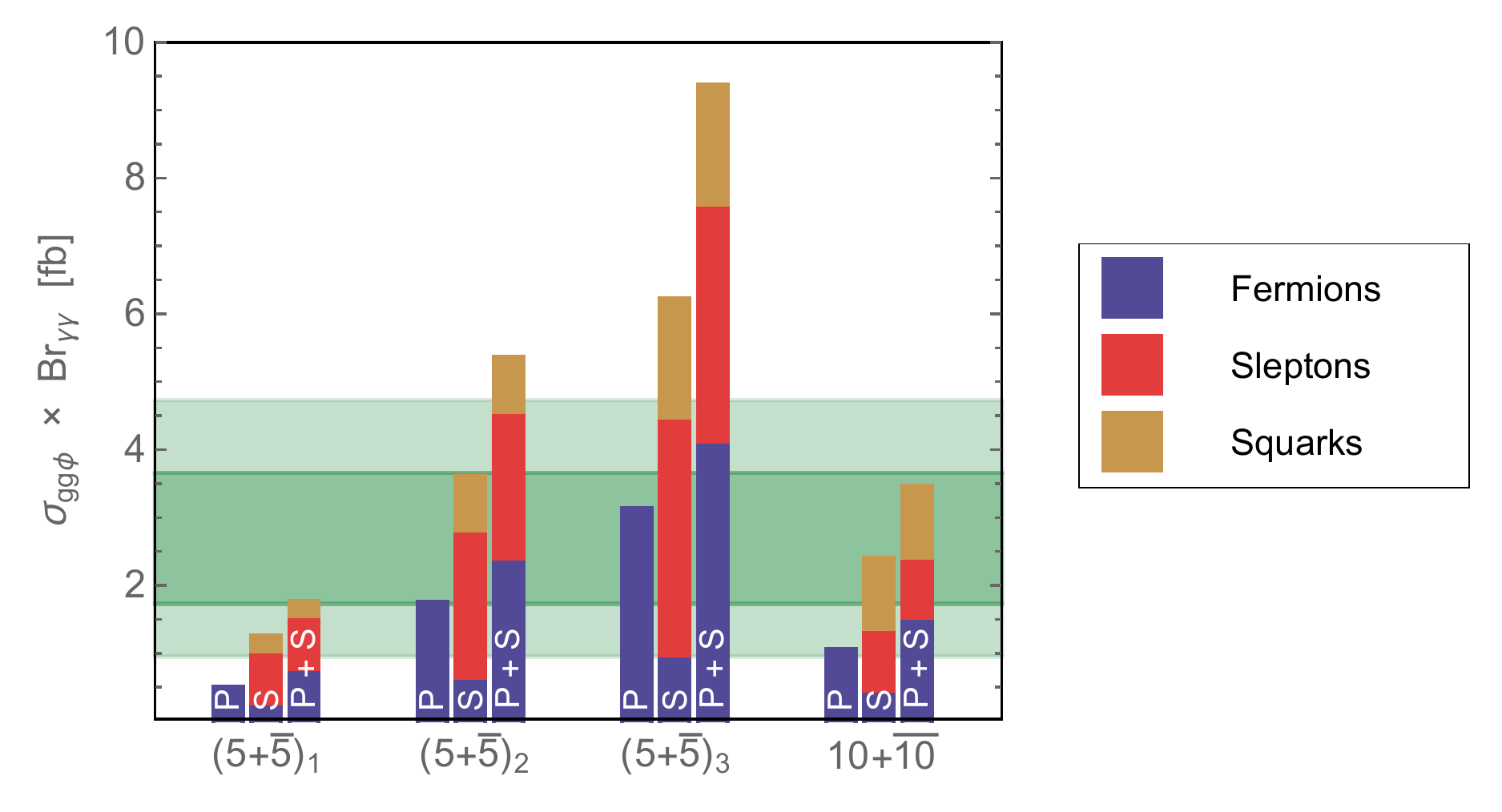}
\end{center}
\caption{Maximal diphoton cross section including fermion and sfermion loops (the individual contributions are shown in different colors). The diphoton resonance can either be identified with the singlet pseudoscalar (P), the scalar (S) or the superposition of both (P+S). The experimentally preferred diphoton cross section ($1$ and $2\,\sigma$ band) is shown in green.}
\label{fig:histo}
\end{figure}

The maximal diphoton cross sections including sfermion loops is shown in figure~\ref{fig:histo}. The diphoton resonance is taken to be either the pseudoscalar singlet, the scalar or the superposition of both. The cross section of the CP even singlet increases dramatically through sfermion effects. The enhancement may reach up to a factor of seven. As a consequence, the observed diphoton signal is now consistent with a single 5-plet of SU(5) if it is caused by the superposition of scalar and pseudoscalar. For all other cases, even the scalar alone can produce a sufficient signal strength. We conclude that the minimal extensions of the MSSM, which are entertained in this work, provide a very attractive explanation to the diphoton excess. In particular, the situation looks very promising if sfermions are not decoupled but reside within the reach of LHC-13.

\section{Predictions for LHC-13}\label{sec:predictions}

\begin{figure}[htp]
\begin{center}
\includegraphics[width=7.65cm]{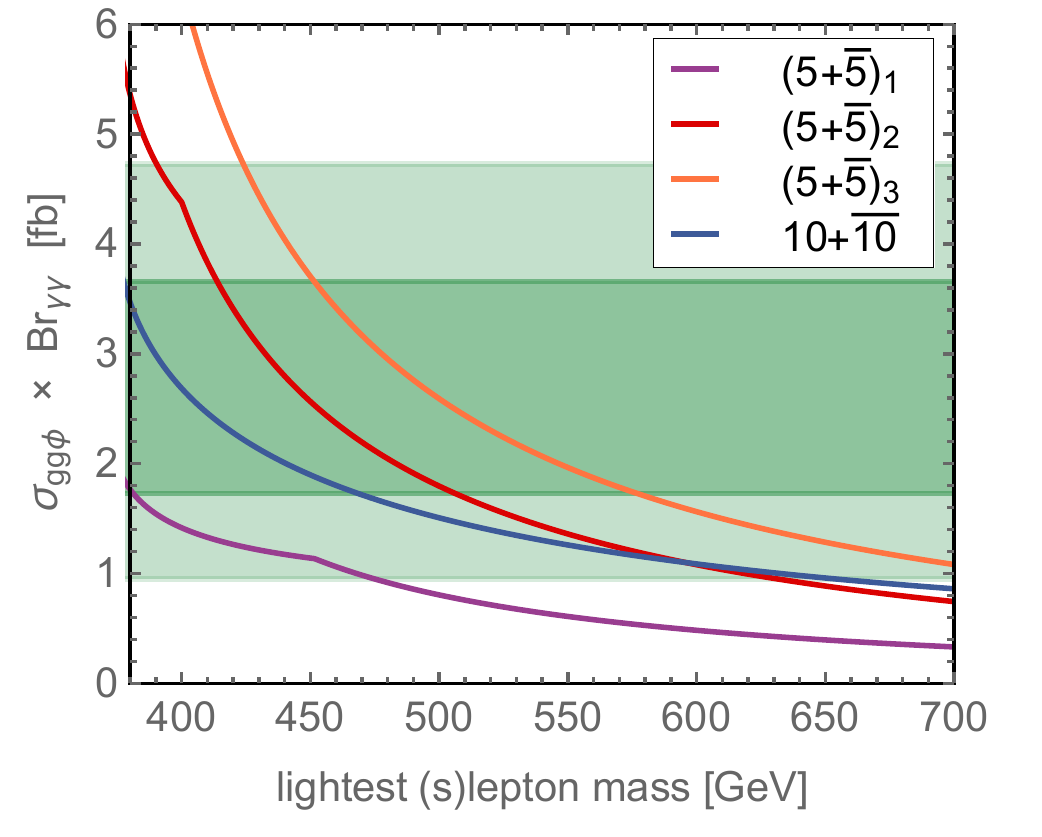}\hspace{3mm}
\includegraphics[width=7.65cm]{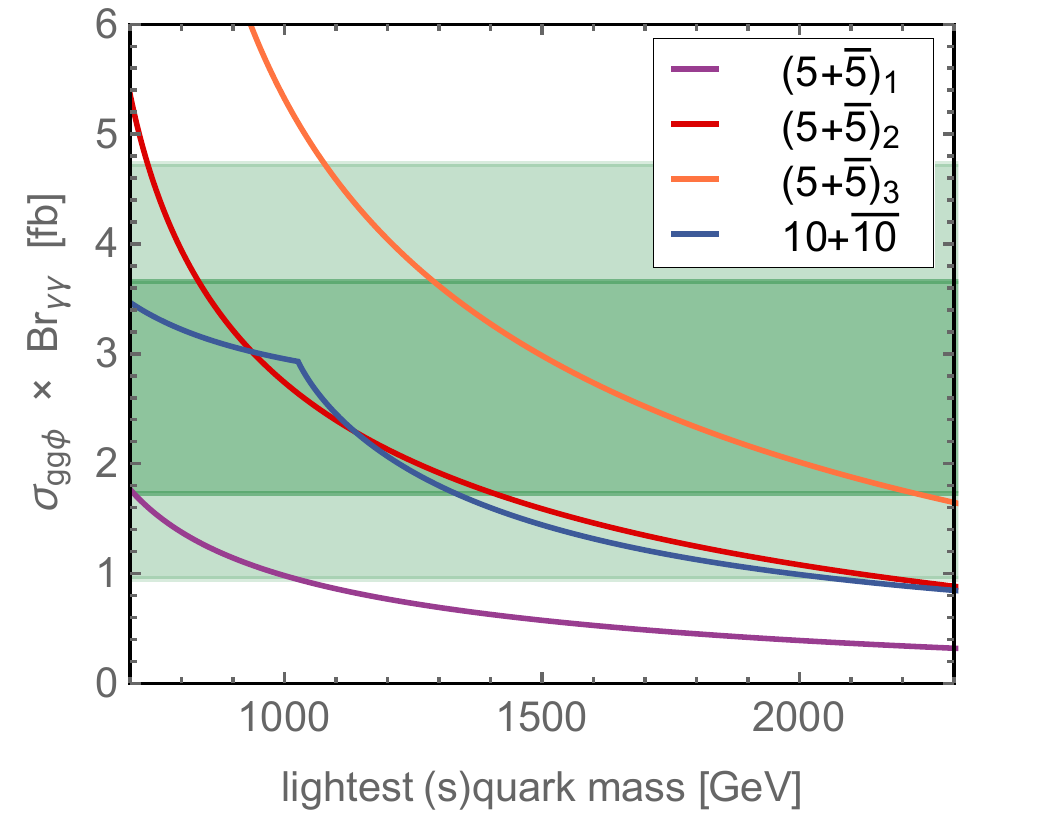}
\end{center}
\caption{Maximal diphoton cross section as a function of the lightest vector (s)lepton mass (left panel) and the lightest vector (s)quark mass (right panel). The experimentally preferred diphoton cross section ($1$ and $2\,\sigma$ band) is shown in green.}
\label{fig:masslimits}
\end{figure}

We now turn to predictions which follow from an explanation to the diphoton excess. The mass spectrum of the vector-like states is severely constrained by requiring sufficient signal strength. In figure~\ref{fig:masslimits} we depict the maximal diphoton cross section as a function of the lightest charged vector particle mass (either a lepton or a slepton) and of the lightest colored vector particle mass (right panel).\footnote{We have identified the resonance with the superposition of scalar and pseudoscalar in order to maximize the diphoton cross section.} In the left panel, we required (s)quark masses of at least $700\:\text{GeV}$, in the right panel slepton masses of at least $380\:\text{GeV}$. In addition, the fermion mass ratios of table~\ref{tab:yukawas} were imposed. Under these boundary conditions, masses and couplings were chosen to maximize the diphoton cross section. 

We observe that consistency with the LHC signal at $1\,\sigma$ imposes an upper limit between $380$ and $580\:\text{GeV}$  on the mass of the lightest vector (s)lepton. The lightest colored vector-like particle cannot be heavier than $0.7-2.2\;\text{TeV}$. The tightest constraints are obtained for the MSSM extended by a single vector-like 5-plet which predicts new charged and colored states `right around the corner'. The weakest limits arise if three $5+\overline{5}$ are introduced. But even in this case, vector-like states are within reach of LHC-13 at high luminosity.

Finally, for a given scenario, we determine the signal strength in the other diboson channels. If the decay of the resonance was completely dominated by one particular loop diagram, the diboson signals would be fixed to the values of table~\ref{tab:simpledecay}. However, we find that generically several vector-like states contribute significantly to the loop decays. Hence, we have to take into account the interference between different loops.

If we assume the best fit cross section from our likelihood analysis, the existing constraints~\cite{Franceschini:2015kwy,Buttazzo:2015txu} from LHC-8 can be translated into\footnote{More restrictive numbers which circulate in the literature have been obtained under the assumption of a larger diphoton cross section.}
\begin{equation}
\frac{\Gamma_{\phi\rightarrow Z\gamma}}{\Gamma_{\phi\rightarrow \gamma\gamma}} < 20\,,\qquad
\frac{\Gamma_{\phi\rightarrow ZZ}}{\Gamma_{\phi\rightarrow\gamma\gamma}} < 21\,,\qquad
\frac{\Gamma_{\phi\rightarrow WW}}{\Gamma_{\phi\rightarrow\gamma\gamma}} < 72\,,\qquad
\frac{\Gamma_{\phi\rightarrow gg}}{\Gamma_{\phi\rightarrow\gamma\gamma}} < 4500\,.
\end{equation}
In case the other diboson final states are only produced via loops (which we assume in our analysis), the constraints are not yet competitive. However, a strong gain in sensitivity is expected at LHC-13. In particular the $Z\gamma$, $ZZ$ and $WW$ channels will provide an important test of diphoton models in the future.

\begin{table}[ht]
\begin{center}
\begin{tabular}{|c|cccc|}
\hline 
\rowcolor{light-gray}&&&&\\[-3mm]
\rowcolor{light-gray}Vector-like Particles& $\Gamma_{\phi\rightarrow Z\gamma}/\Gamma_{\phi\rightarrow \gamma\gamma}$ & $\Gamma_{\phi\rightarrow ZZ}/\Gamma_{\phi\rightarrow \gamma\gamma}$ &$\Gamma_{\phi\rightarrow WW}/\Gamma_{\phi\rightarrow \gamma\gamma}$& $\Gamma_{\phi\rightarrow gg}/\Gamma_{\phi\rightarrow \gamma\gamma}$\\[1mm]
\hline
&&&&\\[-3mm]
$(5+\overline{5})_1$ & $0.4$  & $2.5$ & $6.7-6.8$ & $104-111$ \\[1mm]
$(5+\overline{5})_2$ & $0.3-0.7$  & $2.1-3.0$ & $5.4-8.3$ & $14-313$ \\[1mm]
$(5+\overline{5})_3$ & $0.2-0.7$  & $2.0-3.1$ & $5.2-8.6$ & $7-307$ \\[1mm]
$10+\overline{10}$ & $0-0.5$  & $0.6-2.3$ & $1.0-6.2$ & $87-306$\\ \hline
\end{tabular}
\end{center}
\caption{Branching Ratios of other diboson channels. The given ranges are obtained by requiring consistency with the diphoton signal at $1\,\sigma$.}
\label{tab:dibosonpredictions}
\end{table}

We have scanned over the parameter space of the vector-like extensions of the MSSM. In table~\ref{tab:dibosonpredictions} we provide the resulting predictions in the diboson channels. We have only included the fraction of parameter space compatible with the diphoton excess at $1\,\sigma$. As can be seen, ratios are predicted rather precisely if the vector-like states are comprised by one $5+\overline{5}$. In this scenario, all parameters are basically fixed by requiring a sufficient diphoton signal. If the vector-like sector consists of more 5-plets or a 10-plet there is considerably more freedom to choose parameters. As a consequence, the diboson decay rates may vary within the given ranges. In the $ZZ$ and $WW$ channels, the predicted signals are only a factor of $10$ below the current sensitivity.

\section{From the GUT Scale to the Weak Scale}

We have demonstrated so far that simple vector-like extensions of the MSSM are capable of explaining the LHC diphoton excess. It remains to be shown that the part of parameter space, in which a sufficient diphoton signal arises, is accessible within grand unified theories. 
Therefore, we have chosen to implement the model with a singlet and one $10+\overline{10}$ into SARAH (version 4.8.1)~\cite{Staub:2008uz,Staub:2013tta} which is interfaced with SPHENO (version 3.3.8)~\cite{Porod:2003um,Porod:2011nf}. We slightly adjusted the SARAH  model file `NMSSM10' provided by~\cite{Staub:2016dxq}. The use of SPHENO allows us to study the full two-loop RGE evolution of masses and couplings from high to low energies.

We impose universal boundary conditions on the 10-plets and on the MSSM sector respectively. This results in the following 14 free parameters to be chosen at the GUT scale: $\{m_{1/2},\,m_0\,,A_0,\, \mu,\,B\mu,\tan\beta\}$ in the MSSM sector, $\{\lambda_{10},\,\mu_{10},\,\,B\mu_{10},\,m_{10},\,A_{10}\}$ in the vector-like sector and $\{\mu_{s},\,\,B\mu_{s},\,m_{s}\}$ in the singlet sector. The letters $\mu$, $B$, $m$ stand for supersymmetric, bilinear and soft masses, $A$ for trilinear couplings. 

In table~\ref{tab:benchmark}, we provide a benchmark point which leads to a diphoton signal consistent with the LHC excess. Scalar and pseudoscalar singlet are split by $35\:\text{GeV}$, i.e.\ a double resonance is obtained. The significance in the diphoton channel reaches $3.8\,\sigma$ for this benchmark point. 

\begin{table}[htp]
\centering
 \begin{tabular}{|l r l l r|}
\hline
\rowcolor{light-gray}&&&&\\[-3mm]
\rowcolor{light-gray}
&&& \hspace{-3.6cm}Boundaries at the GUT scale& \\[1mm]
\hline
&&&&\\[-3mm]
$m_{1/2}$               & $2200\:\text{GeV}$           &\qquad\qquad\qquad&  $\lambda_{10}$                     &1.0 \\
$m_0$                   &$1000\:\text{GeV}$              &&  $\mu_{10}$                         & $340\:\text{GeV}$  \\
$A_0$                   &0               &&  $B\mu_{10}$                        &  $-1.97\cdot 10^7\:\text{GeV}$  \\
$\mu$                   &  $800\:\text{GeV}$          & &  $m_{10}$                           &  $2060\:\text{GeV}$ \\
$B\mu$                           &  $6\cdot 10^5\:\text{GeV}$    &       &$A_{10}$                        & 0\\
$\tan\beta$              &       10          &       &         $\mu_s$& $2.54\cdot 10^4\:\text{GeV}$ \\
$m_s$                      &       $9520\:\text{GeV}$       & &           $B\mu_s$ & $-1.51\cdot 10^8\:\text{GeV}$ \\
&&&&\\[-4mm]
\hline
\rowcolor{light-gray}&&&&\\[-3mm]
\rowcolor{light-gray}
&&& \hspace{-2.7cm}MSSM Spectrum&\\[1mm]
\hline
&&&&\\[-3mm]
$m_{\widetilde{B}}$       &  $337\:\text{GeV}$  &&  $m_{h}$ & $127\:\text{GeV}$      \\
$m_{\widetilde{W}}$       &  $594\:\text{GeV}$  &&  $m_{H},\,m_a$ & $3.02\:\text{TeV}$    \\
$m_{\widetilde{g}}$       &  $1.70\:\text{TeV}$  && $m_\text{squarks}$  &  $1.7-3.0\:\text{TeV}$ \\
$m_{\widetilde{h}}$       &   $1.01\:\text{TeV}$      &&   $m_\text{sleptons}$ &  $1.2-1.5\:\text{TeV}$   \\
&&&&\\[-3mm]
\hline
\rowcolor{light-gray}&&&&\\[-3mm]
\rowcolor{light-gray}
&&& \hspace{-4.2cm}Vector-like Spectrum and Couplings&  \\[1mm]
\hline
&&&&\\[-3.5mm]
$\mu_Q$  &  $1.3\:\text{TeV}$  &&  $\lambda_Q$  &  $0.85$    \\
$\mu_U$  &  $1.0\:\text{TeV}$  &&  $\lambda_U$  &  $0.68$    \\
$\mu_E$  &  $376\:\text{GeV}$  &&  $\lambda_E$  &  $0.27$    \\
$m_{\widetilde{Q}_1}$  &  $700\:\text{GeV}$  &&  $T_{Q_1}$        &   $1.47\:\text{TeV}$ \\
$m_{\widetilde{Q}_2}$  &  $3.1\:\text{TeV}$  &&  $T_{Q_2}$        &  $1.31\:\text{TeV}$ \\
$m_{\widetilde{U}_1}$ & $1.4\:\text{TeV}$   && $T_{U_1}$        &    $1.13\:\text{TeV}$ \\
$m_{\widetilde{U}_2}$ & $3.0\:\text{TeV}$   && $T_{U_2}$        &    $0.67\:\text{TeV}$ \\
$m_{\widetilde{E}_1}$ & $375\:\text{GeV}$    && $T_{E_1}$        & $0.61\:\text{TeV}$ \\
$m_{\widetilde{E}_2}$ & $534\:\text{GeV}$    && $T_{E_2}$        & $-0.34\:\text{TeV}$ \\
&&&&\\[-4mm]
\hline
\rowcolor{light-gray}&&&&\\[-3mm]
\rowcolor{light-gray}
&&& \hspace{-2.65cm}Singlet Spectrum& \\[1mm]
\hline
&&&&\\[-3mm]
$m_{h_s}$  &  $744\:\text{GeV}$  &&  \multirow{2}{*}{$m_{\widetilde{s}}$ }& \multirow{2}{*}{$1.37\:\text{TeV}$}    \\
$m_{a_s}$  &  $709\:\text{GeV}$  &&     & \\
&&&&\\[-4mm]
\hline
\rowcolor{light-gray}&&&&\\[-3mm]
\rowcolor{light-gray}
&&& \hspace{-2.55cm}Diphoton Signal & \\[1mm]
\hline
&&&&\\[-3mm]
$\sigma_{gg h_s} \times \text{Br}_{\gamma\gamma}$&  $1.79\:\text{fb}$ &&  \multirow{2}{*}{Significance } &  \multirow{2}{*}{$3.8\,\sigma$}  \\
$\sigma_{gg a_s} \times \text{Br}_{\gamma\gamma}$&  $1.09\:\text{fb}$ &&               &                \\
&&&&\\[-4mm]
\hline
\rowcolor{light-gray}&&&&\\[-3mm]
\rowcolor{light-gray}
&&& \hspace{-3.3cm}Other Diboson Channels & \\[1mm]
\hline
&&&&\\[-3mm]
$\Gamma_{h_s\rightarrow Z\gamma}\,/\,\Gamma_{h_s\rightarrow\gamma\gamma} \!\!\!\!\!\!\!$& 0.1 &&  $\Gamma_{a_s\rightarrow Z\gamma}\,/\,\Gamma_{a_s\rightarrow\gamma\gamma} \!\!\!\!\!\!\!$& 0.1\\
$\Gamma_{h_s\rightarrow ZZ}\,/\,\Gamma_{h_s\rightarrow\gamma\gamma} \!\!\!\!\!\!\!$& 2.3 && $\Gamma_{a_s\rightarrow ZZ}\,/\,\Gamma_{a_s\rightarrow\gamma\gamma} \!\!\!\!\!\!\!$& 1.5   \\
$\Gamma_{h_s\rightarrow WW}\,/\,\Gamma_{h_s\rightarrow\gamma\gamma} \!\!\!\!\!\!\!$& 5.3 &&  $\Gamma_{a_s\rightarrow WW}\,/\,\Gamma_{a_s\rightarrow\gamma\gamma} \!\!\!\!\!\!\!$& 3.6   \\
$\Gamma_{h_s\rightarrow gg}\,/\,\Gamma_{h_s\rightarrow\gamma\gamma} \!\!\!\!\!\!\!$& 205 && $\Gamma_{a_s\rightarrow gg}\,/\,\Gamma_{a_s\rightarrow\gamma\gamma} \!\!\!\!\!\!\!$& 195   \\
\hline 
 \end{tabular}
\caption{Benchmark Point in the MSSM extended by a singlet and a vector-like 10-plet of SU(5). GUT scale boundaries and the resulting particle spectrum, diphoton and diboson signals are shown.}
\label{tab:benchmark}
\end{table}

As can be seen, fermion couplings are perturbative and scalar couplings fulfill the constraints imposed by vacuum stability (cf. table~\ref{tab:trilinears}). Soft parameters have been chosen in the TeV range, only in the singlet sector they are of $\mathcal{O}(10\:\text{TeV})$. There is no fine-tuning associated with this choice as singlet mass terms experience a focus point behavior in the RGE running. At the low scale $\mu_s(\text{TeV}) \ll \mu_s(M_{\text{GUT}})$ and $m_s(\text{TeV}) \ll m_s(M_\text{GUT})$ as a consequence of the large couplings between singlet and vector-like states. Our benchmark example shows that the diphoton signal can indeed be realized via realistic GUT scale boundary conditions.

Turning to the other diboson decay channels, the expected branching ratios are not yet accessible at the LHC. However, if the current sensitivity is increased by one order of magnitude, the predictions in the $ZZ$ and $WW$ channels can be tested.

\section{Conclusion}
Motivated by the diphoton excess at $750\:\text{GeV}$ we have considered a generalization of the MSSM with an additional singlet and vector-like states. Such a setup occurs frequently in string theory models, although no specific arguments are known that point to the particular mass of $750\:\text{GeV}$. The vector-like states tend to be heavy, but some pairs could be protected by specific R-symmetries in an analog way as a Higgs doublet pair is kept light~\cite{Kappl:2008ie}. 

In the present paper, we have concentrated on supersymmetric grand unified theories as a generalization of previous studies performed in~\cite{Hall:2015xds,Tang:2015eko,Dutta:2016jqn}. The novel observation presented here is the potentially decisive role of sfermions in the loop which enhance the fermion contribution to the diphoton signal considered in~\cite{Hall:2015xds} by a factor up to seven. The MSSM completed with a single vector-like 5-plet or 10-plet of SU(5) would be sufficient to explain the excess as seen in the ATLAS and CMS data. In figure~\ref{fig:histo} we present the maximal diphoton cross section in a given vector-like extension requiring
\begin{enumerate}[(i)]
\item perturbative unification of the gauge couplings at $M_\text{GUT}$,
\item the absence of Landau poles in the RGE running of fermion couplings,
\item scalar couplings consistent with a (meta)stable electroweak vacuum.
\end{enumerate}
Supersymmetry breaking can induce a mass splitting between the scalar and pseudoscalar singlet in the vicinity of 750 GeV. We have proven that the corresponding double resonance fits the diphoton spectra as good as a broad resonance.

Apart from the diphoton signal we identified other diboson signals like $Z\gamma$, $ZZ$ and $WW$ which can be deduced in the different extensions of the MSSM with vector-like states as shown in table~\ref{tab:dibosonpredictions}. These specific predictions of the scheme will provide further tests in case the diphoton signal would be confirmed. In table~\ref{tab:benchmark} we specify a complete benchmark model with realistic GUT scale boundary conditions which accommodates the observed diphoton signal.

\section*{Acknowledgments}
We would like to thank Manuel Krauss and Toby Opferkuch for lively discussions and their very helpful comments on the model implementation into SARAH. Further, we thank Florian Staub for correspondence on vacuum stability and Rolf Kappl, Michael Ratz, Patrick Vaudrevange for discussions on the UV embedding of the diphoton excess. This work has been supported by the German Science Foundation (DFG) within the SFB-Transregio TR33 ``The Dark Universe''.

\bibliography{gut}
\bibliographystyle{els}

\end{document}